\documentclass[prb,twocolumn,showpacs,amsmath,amssymb]{revtex4}
\usepackage{graphicx}% Include figure files
\usepackage{dcolumn}% Align table columns on decimal point
\usepackage{bm}% bold math

\begin{document}

\preprint{APS/123-QED}

\title{Drastic annealing effects in transport properties of\\
single crystals of the YbNi$_2$B$_2$C heavy fermion system}% Force line breaks with \\

\author{M. A. Avila}
 \email{avila@ameslab.gov}
\author{S. L. Bud'ko}%
\author{P. C. Canfield}%
\affiliation{%
Ames Laboratory and Department of Physics and Astronomy\\
Iowa State University, Ames, IA 50011
}%

\date{\today}% It is always \today, today,

\begin{abstract}
We report temperature dependent resistivity, specific heat,
magnetic susceptibility and thermoelectric power measurements made
on the heavy fermion system YbNi$_2$B$_2$C, for both as grown and
annealed single crystals. Our results demonstrate a significant
variation in the temperature dependent electrical resistivity and
thermoelectric power between as grown crystals and crystals that
have undergone optimal (150 hour, $950^{\circ}C$) annealing,
whereas the thermodynamic properties:  ($c_p(T)$ and  $\chi (T)$)
remain almost unchanged. We interpret these results in terms of
redistributions of local Kondo temperatures associated with
ligandal disorder for a small ($\sim 1\%$) fraction of the Yb
sites.
\end{abstract}

% PACS, the Physics and Astronomy Classification Scheme.


\maketitle

%\section{\label{sec:intro}Introduction}% force lowercase using \lowercase{text}

Among the rich variety of physical phenomena displayed by the
RNi$_2$B$_2$C nickel boro-carbide family\cite{nag94a,cav94b,
can97a,can98a}, YbNi$_2$B$_2$C (Yb-1221) is unique so far in its
behavior as a heavy fermion system\cite{yat96a,dha96a} with an
electronic specific heat coefficient, $\gamma\approx500 mJ/mol
K^2$, and a Kondo temperature, $T_K\approx10K$: a temperature
scale that is conveniently isolated from other characteristic
temperatures such as superconducting or magnetic condensates
($T_c, T_N < 0.03K$, if they exist at all), and crystal electric
field splitting ($T_{cef}\approx 100K$)\cite{gra96a,ram00a}. The
heavy fermion behavior is understood as a consequence of the fact
that the ytterbium $4f$ levels easily hybridize with the
electronic conduction band levels near the Fermi surface. Such
heavy electron systems often display many unusual physical
properties\cite{ste84b}, one common example being disorder-related
variation in the behavior of different samples of the same
compound\cite{fra78a,ste84b}.

Recently the effects of annealing on the electrical resistivity of
R-1221 single crystals (R = Y, Gd-Lu) were studied\cite{mia02a},
and once again Yb-1221 showed anomalous properties in comparison
to other members of the series. Whereas annealing of crystals with
R = Y, Lu, Tm-Gd simply leads only to changes in the residual
resistivity, radical changes in $R(T)$ behavior were observed for
Yb-1221 over the full ($2K - 300K$) temperature range. The change
in $R(T)$ was so dramatic that comparison of normalized resistance
curves $R(T)/R(300)$ became highly questionable.

To overcome this problem and better understand what is happening
to the Yb-1221 crystals as they go through annealing processes, in
the present work careful resistivity measurements were made on
Yb-1221 crystals cut into well-defined geometries, such that no
normalization is necessary and direct comparison of $\rho(T)$ for
different samples and annealing procedures is possible. The
results, in combination with measurements of specific heat,
magnetic susceptibility and thermoelectric power, allow us to
propose a mechanism in which optimal annealing is essentially
affecting the ligandal disorder associated with a small fraction
($\sim1\%$) of the Yb sites. But given that Yb is a hybridizing
rare earth ion, it is far more sensitive to such disorder than its
non-hybridizing neighbors.  Yb-1221 then presents a model heavy
fermion system that allows for the study of how small changes in
disorder can lead to dramatic changes in transport properties.

%\section{Experimental Details}

Single crystals of YbNi$_2$B$_2$C were grown from Ni$_2$B flux at
high temperatures as described elsewhere\cite{can98a,yat96a}.
Single crystal plates ($4\times4\times0.5 mm^3$ typical
dimensions) were selected, polished and cut into bars (typically
$1-2 mm$ in length and $0.1 mm^2$ cross section) with the length
along the [100] direction using a wire saw.  Once cut the samples
were cleaned with toluene and methanol, placed in a Ta foil
envelope and then placed into the quartz insert of a high
vacuum-annealing furnace. The insert was continuously pumped down
to a pressure of less than $10^{-6} Torr$ during the annealing
schedule which consisted of a one hour dwell at $200^{\circ}C$ to
outgas, followed by the chosen number of hours anneal at
$950^{\circ}C$, followed by a 3-4 hour cool down to room
temperature. Once the schedule was completed, the insert was
brought to atmospheric pressure and the samples removed.  Each
sample was annealed only once, and for the magnetization and
specific heat data the same sample was used to collect the
un-annealed and annealed data sets.

Electrical resistance measurements were performed on Quantum
Design MPMS systems operated in external device control (EDC)
mode, in conjunction with Linear Research LR400/LR700 four-probe
ac resistance bridges, allowing measurements down to $1.8K$. The
electrical contacts were placed on the samples in standard 4-probe
linear geometry, using Pt wires attached to a sample surface with
Epotek H20E silver epoxy. For each slab, the sample weight and
dimensions were carefully measured and an evaluation of the sample
densities ($8.3\pm 0.3 g/cm^3$) was used to estimate an upper
limit of $\pm15\%$ for the overall geometry-related errors in
calculating resistivity $\rho=RA/l$. Susceptibility measurements
were made on these same MPMS systems, using their standard DC
magnetization operating modes, and for the sample oriented in both
$H||c$ and $H||a$ directions.  Specific heat measurements were
made on a Quantum Design PPMS system containing both specific heat
and $^3$He options, allowing measurements down to $0.35K$.

Thermoelectric power measurements were made on a commercial MMR
Technologies Seebeck System, which uses a compressed N$_2$ gas
based Joule-Thompson refrigerator allowing temperature control
down to $90K$. The sample Seebeck coefficient $S(T)$ is obtained
by comparison with a reference constantan wire.

\begin{figure}[htb]
\includegraphics[angle=0,width=88mm]{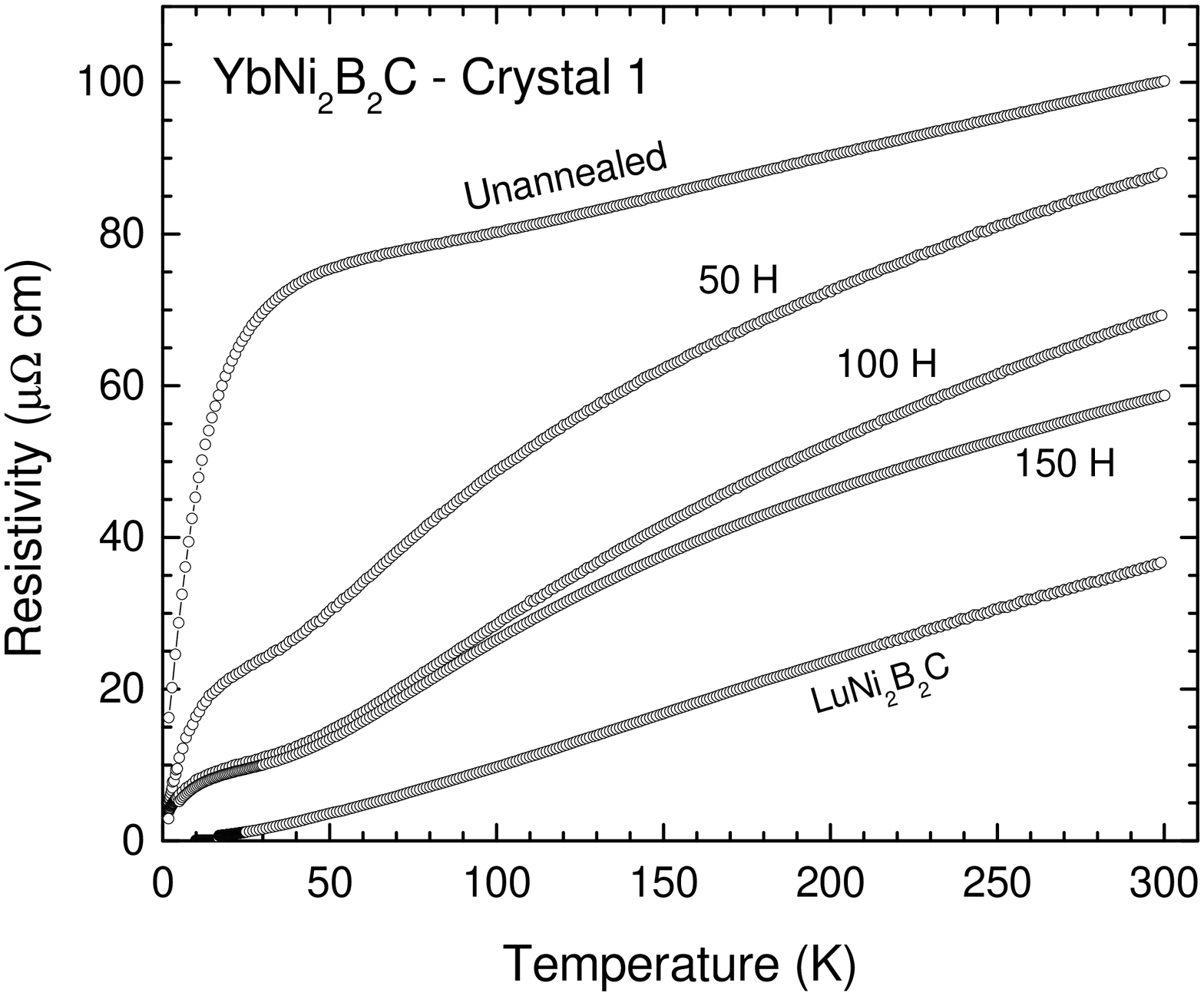}% Here is how to import EPS art
\caption{\label{fig1} Temperature dependence of the electrical
resistivity of several pieces of Yb-1221 cut from crystal 1, for
different annealing times at $950^{\circ}C$. A measurement on
unannealed Lu-1221 is also included for comparison.}
\end{figure}

%\section{Results and Discussion}

The temperature dependence of the electrical resistivity for four
bars cut from a single crystal (crystal 1) is shown in
fig.~\ref{fig1}.  There is a substantial change in the temperature
dependence as well as a reduction in the magnitude of the
resistivity as the annealing time is increased.  For increasing
annealing times the resistivity of Yb-1221 approaches that of
LuNi$_2$B$_2$C (shown as the lowest resistivity curve in
fig.~\ref{fig1})\cite{che98a}.

\begin{figure}[htb]
\includegraphics[angle=0,width=88mm]{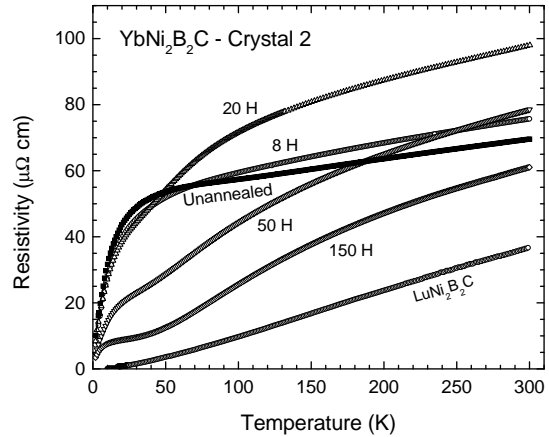}% Here is how to import EPS art
\caption{\label{fig2} Temperature dependence of the electrical
resistivity of several pieces of Yb-1221 cut from crystal 2. Note
that for shorter annealing times the higher temperature
resistivity increases.  For longer annealing times the resistivity
data is quantitatively similar to the data shown in
figure~\ref{fig1}.}
\end{figure}

Figure~\ref{fig2} presents a similar set of annealing data from a
second crystal (crystal 2). In this case several shorter time
anneals were performed in addition to the 50 hour and 150 hour
anneals. Whereas the longer annealing times lead to qualitatively
similar resistivity curves to those shown in fig.~\ref{fig1}, the
shorter annealing times manifest a different behavior:  an
increase in the higher temperature resistivity followed by a
decrease in the lower temperature resistivity.  Qualitatively
similar behaviors were found for several other crystals.  For low
temperatures, resistivity is monotonically reduced with increasing
annealing time, whereas for temperature nearer to room temperature
there can be an initial rise in resistivity followed by an
ultimate decrease in resistivity for increasing annealing times.
Ultimately, for annealing times on the order of 150 hours, the
resistivity is increasingly approaching that of LuNi$_2$B$_2$C.

The temperature dependent resistivity for 150 hour anneals
(figs.~\ref{fig1} and~\ref{fig2}) are consistent with a Yb-based
intermetallic that has $T_K\approx 10K$.  There is an initial
decrease in resistivity with decreasing temperature.  The broad
shoulder near $200 K$ is very likely associated with the thermal
depopulation of the upper CEF levels.  There is a beginning of a
resistance minimum near $30 K$, followed by a sharp decrease in
resistivity associated with coherence as the sample is cooled
through $T_K$. These data are almost precisely what would be
expected from a text book example of a low $T_K$ intermetallic.

\begin{figure}[htb]
\includegraphics[angle=0,width=88mm]{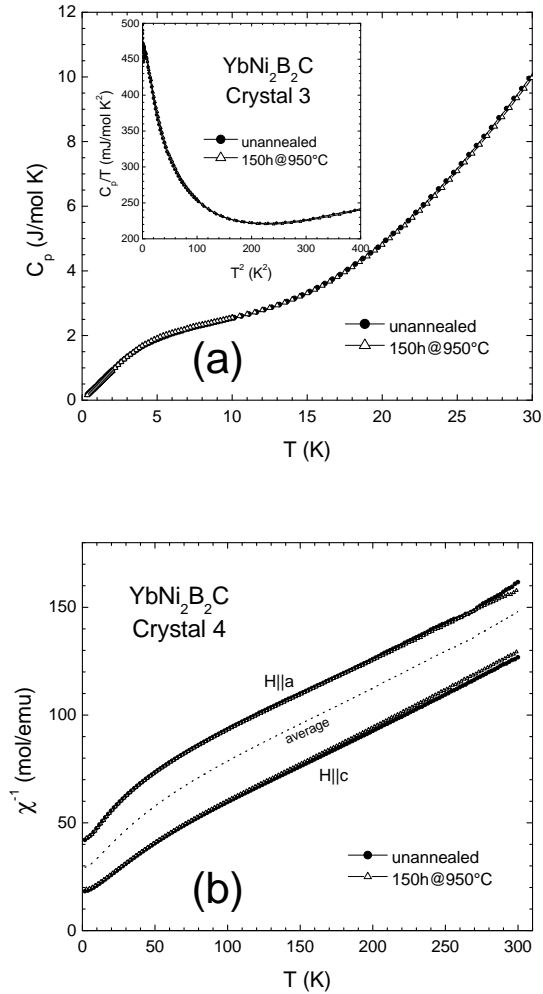}% Here is how to import EPS art
\caption{\label{fig3} (a) Specific heat of Yb-1221 crystal 3,
before and after annealing at $950^{\circ}C$ for 150 hours. The
inset shows the $T^2$ dependence of the linear term $c_p/T$ at the
lowest temperatures. (b) Inverse susceptibility of Yb-1221 crystal
4 for both orientations $H||c$ and $H||a$, before and after
annealing at $950^{\circ}C$ for 150 hours. Annealing does not
significantly change the crystal's thermodynamic properties.}
\end{figure}

In contrast with the results of resistivity measurements, 150 hour
annealing at $950^{\circ}C$ causes only minor changes (if any) in
the material's bulk thermodynamic properties. In fig.~\ref{fig3}
we show measurements of the temperature dependence of (a) specific
heat and (b) inverse magnetic susceptibility performed on two
pieces cut from crystals 3 and 4 respectively, before and after
they were annealed for $150h@950^{\circ}C$. The general features
of these curves are consistent with previously reported
measurements on unannealed Yb-1221 crystals\cite{yat96a}. The fact
that the bulk thermodynamic properties do not change, within
experimental limits, gives support to the hypothesis that
annealing is affecting only a very small fraction of the sample,
and also helps to rule out any possibility of major structural
changes such as phase separation, decomposition, corrosion,
oxidization or other such degradations to the crystalline lattice.

Based on the data presented in figures 1-3 as well as previous
work on the non-hybridizing members or the RNi$_2$B$_2$C series
(e.g. R = Y, Lu, Tm-Gd)\cite{mia02a} it appears that annealing
single crystals of RNi$_2$B$_2$C can cause significant changes in
the electrical resistivity.  For the samples of LuNi$_2$B$_2$C or
YNi$_2$B$_2$C (as well as the local moment members) this increased
perfection simply leads to a reduction of the residual
resisitivity.  On the other hand for the case of YbNi$_2$B$_2$C,
disorder around the hybridizing Yb ion can very easily lead to a
change of the local Kondo temperature.  This observation leads to
a simple hypothesis:  a small number of Yb sites have a
distribution of Kondo temperatures that can have values ranging
from 10 K or lower up to above room temperature. This distribution
in $T_K$ values is associated with a distribution of Yb-ligandal
environments.  With annealing, this disorder is reduced and a
diminishingly small number of Yb sites have Kondo temperatures
different from $T_K\approx 10 K$.

The first question that can be simply addressed is: what
percentage of Yb sites would need to have Kondo temperatures above
room temperature to account for the $\sim40\mu\Omega cm$ excess
resistivity found near room temperature in the unannealed samples?
This can be estimated by using the expression for the increase in
resistivity associated with a Kondo impurity for temperatures
below the Kondo temperature associated with the
impurity\cite{tho85a}:

\begin{equation}
\Delta\rho_{max}=\frac{\hbar}{e^2}\frac{4\pi c}{pk_F}(2l+1)
\label{eq:rho}
\end{equation}

where $c$ is the concentration of Yb ``impurities'', $k_{F}$ is
the Fermi momentum, $p$ is the number of electrons per atom, and
$l$ is the ytterbium $4f$ angular momentum. Using $l=3$, $p=3$ and
$k_F\approx3.7\times10^7 cm^{-1}$
[refs.~\onlinecite{kog97a,dug99a}] in this equation, we estimate
$\Delta\rho_{max}=30\mu\Omega cm$ for every 1\% concentration of
disordered Yb sites with $T_K$ greater than room temperature. This
means that only a few percent of the Yb ions need to have Kondo
temperatures different from the expected value of $T_K\approx10K$,
to produce a significant increase in resistivity that is seen in
figs.~\ref{fig1} and ~\ref{fig2}. It should be noted that the
above calculation is at best semi-quantitative (i.e. should be
trusted to only within factors of two), but it does demonstrate
that our working hypothesis is a viable one.

The evolution of the temperature dependent resistivity can be
understood then by simply invoking a reduction of ligandal
disorder with annealing.  For long time anneals the number of Yb
ions that have $T_K$ values higher than $10 K$ has been reduced to
a negligible amount.  This leads to the nearly classical $\rho(T)$
curves seen in figs.~\ref{fig1} and~\ref{fig2}. For smaller
annealing times the higher temperature resistivity of the sample
can either increase or decrease, depending upon the nature
(quantity and type) of disorder in the as-grown sample.  As the
ligandal disorder is annealed away there is no reason to assume
that intermediate states (of crystallographic order) will have
$T_K$ values that monotonically approach $10 K$:  e.g. as seen in
crystal 2 (fig.~\ref{fig2}) small annealing times appear to be
populating configurations with higher $T_K$ values. We have
performed annealing studies on five different groups of samples
cut from crystals from different batches. In every case the long
time anneals ($\sim150$ hours) approach the resistivity data shown
in figs.~\ref{fig1} and~\ref{fig2}. The behavior illustrated by
figs.~\ref{fig1} and~\ref{fig2} appears to be a generic behavior
for Yb-1221 samples prepared in this manner.

If indeed the changes in $\rho(T)$ are due to changes in ligandal
disorder in a small fraction of the Yb-sites, then we would
anticipate that any bulk, thermodynamic property of the material
should not be significantly affected by the annealing process. For
$c_p(T)$, the change in the number of $T_K\approx 10 K$ Yb sites
(on the order of 98\% to 100\%) that occurs with annealing is so
small that it is not resolvable in our measurements
(fig.~\ref{fig3}a) on the same piece of sample (as grown and
annealed). The measured curves are essentially indistinguishable
for $0.35K\leqslant T\leqslant 30K$. For $\chi(T)$, annealing
effects are equally indistinguishable. The vast majority of the Yb
sites are acting as Yb$^{+3}$ at high temperatures and, again, a
change of a percent or two is not distinguishable in our
measurements.

\begin{figure}[htb]
\includegraphics[angle=0,width=88mm]{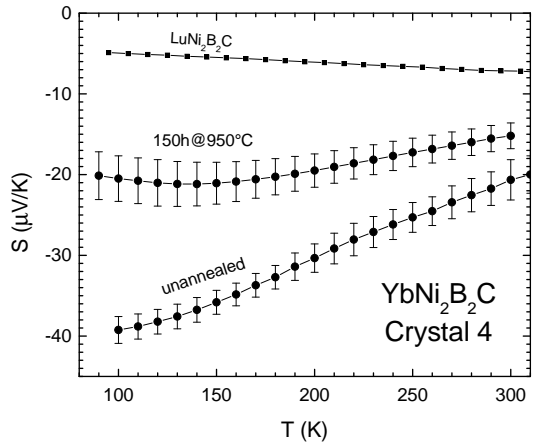}% Here is how to import EPS art
\caption{\label{fig4} Absolute thermopower of several samples of
Yb-1221 cut from crystal 4 (the unannealed data manifold is the
average of data from three samples, and the annealed data manifold
is the average of data from two samples that were annealed at
$950^{\circ}C$ for 150 hours). A measurement on unannealed Lu-1221
is also included for comparison.}
\end{figure}

Finally, since the assumed disorder affects the DOS profile near
the Fermi surface, one should expect significant changes in the
material's thermoelectric power. Figure~\ref{fig4} shows the
temperature dependence of the Seebeck coefficient $S(T)$ obtained
by measuring several pieces cut out of Yb-1221 crystal 4. Three
pieces were measured after cutting, and two other pieces were
measured after annealing for $150h@950^{\circ}C$, and the results
for these two groups of data were averaged. An unannealed Lu-1221
sample was also measured for comparison. Both the unannnealed
Yb-1221 and Lu-1221 data agree very well with previously reported
measurements on these compounds\cite{rat99a}, although there is a
sample-to-sample scatter (as indicated by the error bars) in
$S(T)$ in our data.

The limited temperature range of our experimental equipment does
not allow us to clearly observe the most prominent feature in such
measurements, which is a large negative peak in $S(T)$ just below
$100K$\cite{rat99a} but the high-temperature data of our samples
is seen to follow the generally expected behavior within the
framework of the Coqblin-Schrieffer
model\cite{zla01a,bha76a,coq69a}, and the disorder-related effects
discussed here. For the unannealed samples, the response should be
a combination of a low $T_K$ Kondo lattice and single-Yb
``impurity'' effects with high $T_K$ values, both of which are
known to greatly increase the absolute values of $S(T)$ for $T\gg
T_K$. Annealing will then have the effect of reducing the impurity
contribution, which in our case results in a reduction of roughly
50\% (from $\sim40$ to $\sim20\mu V/K$) in the magnitude of the
thermopower response near the peak, whose position also seems to
shift to slightly higher temperatures. The resulting curves for
annealed Yb-1221 samples move towards the Lu-1221 reference, but
still maintain the enhanced contribution coming mostly from the
Kondo Lattice with CEF splitting\cite{bha76a}.

%\section{Conclusion}

In conclusion, we have shown how annealing of single crystal
YbNi$_2$B$_2$C leads to dramatic changes in its transport
behavior. These changes can be understood as a consequence of the
effects of ligandal disorder that leads to locally large values of
$T_K$ associated with a small fraction of local Yb sites.  This
small number of high $T_K$ sites leads to large enhancements in
electrical resistivity and thermoelectric power that are very
sensitive to the amount of time the crystal is annealed. Increased
annealing times reduce the number of perturbed Yb sites and lead
to more classical, and simply understood, temperature dependencies
of the transport properties. To this extent, YbNi$_2$B$_2$C
continues to be a model heavy fermion compound: not only from the
fact that other characteristic energy scales are well separated
from $T_K$ in this material\cite{yat96a}, but also because it
allows for more detailed investigations of exactly how different
disturbances in the vicinity of Yb sites affect the hybridization
of its $4f$ level with conduction states and the hybridized Kondo
lattice. This is a near ubiquitous problem in heavy fermion
systems and will hopefully be addressed more fully by future
diffraction measurements aimed at determining the exact nature of
the hypothesized ligandal disorder.

\begin{acknowledgments}
We are thankful to F. Steglich for suggesting the experiments on
thermoelectric power. Ames Laboratory is operated for the US
Department of Energy by Iowa State University under Contract No.
W-7405-Eng-82. This work was supported by the Director for Energy
Research, Office of Basic Energy Sciences.
\end{acknowledgments}

%\bibliography{ameslab,boro-carbides,condmat}% Produces the bibliography via BibTeX.

\end{document}